\documentclass[showpacs,preprintnumbers,amsmath,amssymb,floatfix]{revtex4-1}

\usepackage{graphicx}
\usepackage{dcolumn}
\usepackage{bm}
\usepackage{amssymb}
\usepackage{epsfig}
\usepackage{color}

\newcommand{\ba}{\begin{eqnarray}}
\newcommand{\ea}{\end{eqnarray}}
\newcommand{\be}{\begin{equation}}
\newcommand{\ee}{\end{equation}}
\newcommand{\bdisplay}{\begin{displaymath}}
\newcommand{\edisplay}{\end{displaymath}}

\newcommand{\eq}[1]{Eq.\,(\ref{#1})}

\makeatletter
\def\eqnarray{\stepcounter{equation}\let\@currentlabel=\theequation
\global\@eqnswtrue
\tabskip\@centering\let\\=\@eqncr
$$\halign to \displaywidth\bgroup\hfil\global\@eqcnt\z@
  $\displaystyle\tabskip\z@{##}$&\global\@eqcnt\@ne
  \hfil$\displaystyle{{}##{}}$\hfil
  &\global\@eqcnt\tw@ $\displaystyle{##}$\hfil
  \tabskip\@centering&\llap{##}\tabskip\z@\cr}

\def\endeqnarray{\@@eqncr\egroup
      \global\advance\c@equation\m@ne$$\global\@ignoretrue}

\def\@yeqncr{\@ifnextchar [{\@xeqncr}{\@xeqncr[5pt]}}
\makeatother

\begin{document}

\title{ Mehler-Fock transforms and retarded radiation Green functions in hyperbolic and spherical spaces}
\author{ Loyal Durand}
\email{ldurand@hep.wisc.edu}
\altaffiliation{Present address: 415 Pearl Court, Aspen, CO 81611}
\affiliation{Department of Physics
University of Wisconsin-Madison
Madison, WI 53706}

\begin{abstract}
We develop the theory of causal radiation Green functions on hyperbolic and hyperspherical spaces using a constructive approach based on generalized Mehler-Fock transforms. This approach focuses for $H^d$ on the kernel of the transformation expressed in terms of hyperbolic angles $\theta$ with $0\leq\theta<\infty$. The kernel provides an explicit representation for the generalized delta distribution which acts as the source term for the radiation, and allows easy implementation of the causality or retardation condition and determination of the Green function. We obtain the corresponding kernel distribution on $S^d$ by analytic continuation of the kernel distribution of the Helmholtz equation on $H^d$, then show that this construction  leads to the proper retarded Green function for the wave equation. That result  is then used to establish the validity of a new generalized Mehler-Fock transformation for $0\leq\theta<\pi$. The present results clarify and extend those obtained recently by Cohl, Dang, and Dunster.
\end{abstract}


\maketitle


\section{Introduction \label{sec:intro} }

In a recent paper \cite{CohlDangDunster}, Cohl, Dang, and Dunster made a very thorough analysis of the scalar Green functions or fundamental solutions for the wave and Helmholtz equations in the hyperbolic and hyperspherical spaces $H^d$ and $S^d$. They obtained a number of strong results on the asymptotic behavior of the functions that appear as solutions to those equations, as well as several theoretical constraints on the Green functions.  They then constructed candidate Green functions from the solutions of those equations, and determined the proper result on $H^d$, with its normalization, from the requirement that it reduce to the known result in  $E^d$ in the Euclidean or flat-space limit. This method failed for the wave equation on $S^d$, where they obtained two candidate Green functions which could not be distinguished by the limiting process. It was also not possible to impose the retardation condition necessary for causal wave propagation directly. It appeared only implicitly through the Sommerfeld radiation condition used in the usual construction of the causal Green function on $E^d$.

In the present paper, we will take a different approach for the case of radiation Green functions on $H^d$ and $S^d$. We will construct the Green functions directly using generalized Mehler-Fock transforms. This makes it simple to impose the requirement of causality, that the functions constructed be retarded Green functions so that no signal can reach a point a geodesic distance $\lvert{ x}\rvert$ away from the source point in time less than $\lvert{x}\rvert/c$, $c$ the wave speed or speed of light. The results for $S^d$ clarify those obtained by  Cohl, Dang, and Dunster  \cite{CohlDangDunster}, and pick one of their candidate solutions with the normalization determined.  

We will present our results in terms of the Gegenbauer or hyperspherical functions which appear naturally as solutions of the wave equation rather than the Legendre and Ferrers functions used by those authors. Our approach has the advantage that it also yields the retarded Green functions associated with the Gegenbauer equation for functions of general order, with $d$ non-integer as encountered in dimensional continuation in quantum field theory.

Our approach is constructive, using the defining relations for the Green function in terms of the  inhomogeneous wave equation with a generalized delta distribution as the source term. We implement the construction using generalized Mehler-Fock transforms, with the combined kernel of the initial integral transform and its inverse treated as a Schwarz distribution. These transforms are natural on $H^d$, and allow a simple construction of the Green function and implementation of the causality condition. Conversely, our construction puts the theory of the transforms in a distribution-related context,  treated in terms of the kernel distributions rather than integral transforms and inverses.

To construct the Green function on $S^d$, we develop a new Mehler-Fock transform applicable for spherical angles $\theta$ with $\cos{\theta}$ on the interval $(-1,1)$.  We derive it initially though a continuation of the kernel of the transform appropriate for the Helmholtz equation on $H^d$, construct the hyperspherical Green function and establish its validity, and then use the results to establish the the validity of the new transform. A more direct derivation would be of interest.

The outline of the paper is as follows. We will first discuss the general background in Sec.~\ref{sec:preliminaries}, introducing our coordinates on $H^d$ and $S^d$ in Sec.~\ref{subsec:coordinates}. We summarize the solution of the wave equation on those spaces and  relevant properties of the Gegenbauer function of the first and second kind which appear in those solutions in Sec.~\ref{subsec:wave equation}, and the conditions for the construction of the scalar Green function in Sec.~\ref{subsec:constructG}. 

We construct the retarded scalar Green function on $H^d$ using a generalized Mehler-Fock transform  in Sec.~\ref{sec:MehlerFock}.  We introduce the transform  we will use in Sec.~\ref{subsec:MehlerFock}, study the properties of the kernel of the transform in Sec.~\ref{subsec:MehlerFork_kernel_Hd}, and use the results to derive a form of the retarded Green function in Sec.~\ref{subsec:G_R}. We then derive the scalar Green function on $H^d$ in Sec.~\ref{subsec:scalar_Greens_func}. The result is unique. It agrees with that of Cohl, Dang, and Dunster \cite{CohlDangDunster}, but was constructed using the retardation condition directly rather  than by requiring agreement with the known  Euclidean limit.

We construct the scalar Green function on $S^d$ in Sec.~\ref{sec:generalizedMehlerFock}. We first construct a generalized Mehler-Fock kernel on $S^d$ in Sec.~\ref{subsec:MehlerFock_Sd} by analytic continuation in the distribution sense from the case of the Helmholtz equation on $H^d$. We use the result in Sec.~\ref{subsec:retardedGonSd} to construct the retarded scalar Green function on $S^d$. We then show directly in Sec.~\ref{subsec:MehlerFock_onSd} that the action of the wave operator on this Green function leads to the proper  kernel for a Mehler-Fock type transform for spherical angles with $0<\theta<\pi$. We present this transform in  two different forms. It appears to be new.


\section{Preliminaries \label{sec:preliminaries}}    


\subsection{Coordinates on $H^d$ and $S^d$ \label{subsec:coordinates}}

$H^d$ and $S^d$ are homogeneous spaces with all points equivalent. We assume there are no boundaries in either case. The scalar Green functions can therefore only depend on the scalar distance $\lvert x-x'\rvert$ between the source point $x'$  and the field point $x$, and not on those points individually. This distance is invariant under the hyperbolic or hyperspherical rotations that move those points in the respective spaces. Causality requires that the Green function vanish for $\lvert x-x'\rvert>ct$ for a signal that originates at $x'$  at time $t=0$. 

The spaces $H^d$ and  $S^d$ will be taken as embedded in $d+1$   dimensional hyperbolic and hyperspherical spaces, with the the hyperboloids and hyperspheres having fixed radius $R$. We will use coordinates $x=(x_0,x_1,\cdots,x_d)$ corresponding to the reductions $H(d,1))/SO(d)$ and $SO(d+1)/SO(d)$ of the groups of symmetry transformations in those spaces, with
\ba
x_0&=&R\cosh{\theta}, \quad x_1=R\sinh{\theta}\cos{\theta_1},\quad x_2=R\sinh{\theta}\sin{\theta_1}\cos{\theta_2},\ \cdots,\nonumber \\
\label{Hd_coord}
x_{d-1}&=&R\sinh{\theta}\sin{\theta_1}\cdots\sin{\theta_{d-1}}\cos{\phi}, \quad x_d=R\sinh{\theta}\sin{\theta_1}\cdots\sin{\theta_{d-1}}\sin{\phi},
\ea
for $H^d$, with $x^2=x_0^2-x_1^2\cdots-x_d^2=R^2$, and
\ba
x_0&=&R\cos{\theta}, \quad x_1=R\sin{\theta}\cos{\theta_1},\quad x_2=R\sin{\theta}\sin{\theta_1}\cos{\theta_2},\ \cdots,\nonumber \\
\label{Sd_coord}
x_{d-1}&=&R\sin{\theta}\sin{\theta_1}\cdots\sin{\theta_{d-1}}\cos{\phi}, \quad x_d=R\sin{\theta}\sin{\theta_1}\cdots\sin{\theta_{d-1}}\sin{\phi},
\ea
for $S^d$, with $x^2=x_0^2+x_1^2\cdots+x_d^2=R^2$. 

The geodesic distances $\lvert x-x'\rvert$ between points on $H^d$ and $S^d$ are given in hyperbolic or hyperspherical geometry simply by $R\Theta$ where $\Theta$ is the hyperbolic (hyperspherical) angle between the points, with $x\cdot x'=R^2\cosh{\Theta}$ on $H^d$ and $x\cdot x'=R^2\cos{\Theta}$ on $S^d$. Because of the homogeneity of the spaces, we can use appropriate rotations to greatly simplify the expressions for $x$ and $x'$. Thus, on either space, we can choose coordinates such that the $3,\,4,\,\cdots,d$ components of both $x$ and $x'$  siultaneously. We can then write the 1 and 2 components of $x$ as $x_1=R\sinh{\theta}\cos{\phi_1}$ and $x_2=R\sinh{\theta}\sin{\phi_1}$ with corresponding expressions for $x'$; $\cosh\Theta $ is then given by
\be
\label{ThetaH}
\cosh{\Theta} =\cosh{\theta}\cosh{\theta'}-\sinh{\theta}\sinh{\theta'}\cos{\varphi} 
\ee
with $\varphi=\phi_1-\phi_2$. Similarly, on $S^d$,
\be
\label{ThetaS}
\cos{\Theta} = \cos{\theta}\cos{\theta'}+\sin{\theta}\sin{\theta'}\cos{\varphi}. 
\ee
The scalar Green functions can depend only  on $\Theta$ in either case. It will further be useful  at some points later to take $\varphi=0$. With this choice of coordinates $x$ and $x'$ both lie along the 1 axis with separation $R\Theta$.


\subsection{Solution of the wave equation \label{subsec:wave equation}}


The wave equation in $d+1$ dimensions is
\be
\label{wave_eq1}
\left(-\bigtriangleup+\frac{1}{c^2}\frac{d^2}{dt^2}\right){\mathfrak f}(x,\,t)=0, \quad x=(x_0,x_1,\cdots,x_d),
\ee
with $\bigtriangleup$ the Laplacian in the chosen coordinates $x$ and $c$ the wave propagation speed. This equation is separable in the coordinates above and the time $t$. Defining the frequency-dependent function $f(x,\omega)$ as
\be
\label{f_omega}
f(x,\omega) = \int_{-\infty}^\infty dt\,{\mathfrak f}(x,t)e^{i\omega t}
\ee
with the inverse
 \be
 \label{f_t}
 {\mathfrak f}(x,t) = \frac{1}{2\pi}\int_{-\infty}^\infty d\omega f(x,\omega) e^{-i\omega t}, 
 \ee
 we have
 \be
 \label{wave_eq2}
\left(-\bigtriangleup-\frac{\omega^2}{c^2}\right)f(x,\omega)=0.
\ee

The frequency-dependent Green function $G(x,x',\omega)$ satisfies the corresponding inhomogeneous equation
\be
\label{Greens_eq}
\left(-\bigtriangleup-\frac{\omega^2}{c^2}\right)G(x,x',\omega) = \delta^{(d+1)}(x-x')
\ee
with $\delta^{(d+1)}$ the expression for the $d+1$-dimensional delta distribution in the chosen coordinates (\cite{CohlDangDunster}, Sec.~4.2). We note that $G(x,x',\omega)$ must itself be a solution of the homogeneous wave equation except in a neighborhood of the singularity.

The  angular components of the solutions of \eq{wave_eq2} can be described in terms of hyperspherical harmonics $Y^{m_1,\cdots,m_d}$, expressible as products of Gegenbauer polynomials (see, e.g., \cite{HTF}, Sec.\ 11.2, and \cite{Vilenkin}, Chap.\ IX). These are solutions of the reduced Laplace equation on $S^{d-1}$, 
\be
\label{Laplace_Y}
-\bigtriangleup_{d-1}(\theta_1,\cdots,\phi) Y^{m_1,m_2,\cdots,m_d}(\theta_1,\cdots,\theta_{d-1},\phi)=m_1(m_1+d-2)Y^{m_1,m_2,\cdots,m_d}(\theta_1,\cdots,\theta_{d-1},\phi)
\ee
in the $d$ angular coordinates, with $\bigtriangleup_{d-1}(\theta_1,\cdots,\phi)=R^2\bigtriangleup_{d-1}(x_1,\cdots,x_d)$. We will relabel $m_1$ as $l$, the angular momentum  associated with the hyperspherical functions, and write the angular functions as $Y_l^M$, $M$ the multi-index $(m_2,\cdots,m_d)$. (In the notation of \cite{HTF}, Sec.\ 11.2, $l=m_0$ and  $M=(m_1,\cdots,m_{d-1})$ in $d+1$ dimensions.)

The remaining radial components satisfy the equations
\be
\label{wave_eq_H}
\left[\left(-\frac{d^2}{d\theta^2}-(d-1)\coth{\theta}\frac{d}{d\theta}+\frac{l(l+d-2)}{\sinh{\theta}^2}\right)-k^2R^2\right]w(\theta)=0
\ee
for $H^d$, and
\be
\label{wave_eq_S}
\left[\left(-\frac{d^2}{d\theta^2}-(d-1)\cot{\theta}\frac{d}{d\theta}+\frac{l(l+d-2)}{\sin{\theta}^2}\right)-k^2R^2\right]w(\theta)=0
\ee
for $S^d$, with $k=\omega/c$ the usual wavenumber. 

In the case $l=0$ which we will need in the case of the scalar Green functions, these equations are equivalent to the Gegenbauer equations
\be
\label{GegenbauerHd}
\left(\frac{d^2}{d\theta^2}+2\alpha\coth{\theta}\frac{d}{d\theta}-(\nu^2-\alpha^2)\right)w_{\nu-\alpha}^\alpha(\cosh{\theta})=0
\ee
for $H^d$ with $\alpha=(d-1)/2$ and $\nu^2-\alpha^2=-k^2R^2$,  and
\be
\label{GegenbauerSd}
\left(\frac{d^2}{d\theta^2}+2\alpha\cot{\theta}\frac{d}{d\theta}+(\nu^2-\alpha^2)\right)w_{\nu-\alpha}^\alpha(\cos{\theta})=0
\ee
for $S^d$ with $\nu^2-\alpha^2=k^2R^2$.

 In these expressions $w_{\nu-\alpha}^\alpha$ is a Gegenbauer function of either the first or second kind, $C_{\nu-\alpha}^\alpha$ or $D_{\nu-\alpha}^\alpha$,
defined as
\ba
\label{Cdefined}
C_{\nu-\alpha}^\alpha (z)&=& \frac{\Gamma(\nu+\alpha)}{\Gamma(\nu-\alpha+1)\Gamma(2\alpha)}\, _2F_1\left(-\nu+\alpha,\nu+\alpha;\alpha+\frac{1}{2};\frac{1-z}{2}\right), \\
D_{\nu-\alpha}^\alpha(z) &=& e^{i\pi\alpha}[2(z-1)]^{-\nu-\alpha}\frac{\Gamma(\nu+\alpha)}{\Gamma(\nu+1)\Gamma(\alpha)} \nonumber \\
\label{Ddefined}
&& \times\,_2F_1\left(\nu+\alpha,\nu+\frac{1}{2};2\nu+1;\frac{2}{1-z}\right),
\ea
 (see \cite{HTF},\ Sec.~3.15; or \cite{szego}, Sec. 4.7).

More generally, for $l\not=0$,
the respective substitutions $w(\theta)=\left(\sinh{\theta}\right)^{-\alpha}v(\theta)$ and $w(\theta)=\left(\sin{\theta}\right)^{-\alpha}v(\theta)$ in Eqs.\ (\ref{wave_eq_H}) and (\ref{wave_eq_S}), again with $\alpha=(d-1)/2$, bring these equations to a form of the Gegenbauer equation  considered by Szeg{\H o} (\cite{szego}, Eq.~4.7.11),
\be
\label{wave_eq_H_2}
\frac{d^2v}{d\theta^2}-\frac{(l+\alpha)(l+\alpha-1)}{\sinh^2\theta}v = \left(-k^2R^2+\alpha^2\right)v
\ee
for $H^d$, and
\be
\label{wave_eq_S_2}
\frac{d^2v}{d\theta^2}-\frac{(l+\alpha)(l+\alpha-1)}{\sin^2\theta}v = \left(-k^2R^2-\alpha^2\right)v
\ee
for $S^d$. The solutions are $v=(\sinh{\theta})^\mu F_{\nu-\mu}^\mu(\cosh{\theta})$, or $v=(\sin{\theta})^\mu F_{\nu-\mu}^\mu(\cos{\theta})$, where $F_{\nu-\mu}^\mu$ is a Gegenbauer function of either the  first or second kind. From \cite{szego}, Eq.~4.7.11,  $\mu(\mu-1)=(l+\alpha)(l+\alpha-1)$ giving $\mu=l+\alpha$ or $\mu=-l-\alpha+1$, while $\nu^2=-k^2R^2+\alpha^2$ giving $\nu=\pm i\sqrt{k^2R^2-\alpha^2}$ for \eq{wave_eq_H_2}. Similarly, $\nu=\pm \sqrt{k^2R^2+\alpha^2}$ for \eq{wave_eq_S_2}, so with $\sigma=\sqrt{k^2R^2-\alpha^2}$, $w=(\sinh{\theta})^{\mu-\alpha}F_{\pm i\sigma-\mu}^\mu(\cosh{\theta})$ for $H^d$.

While there are nominally eight solutions given the two forms each for $\mu$ and $\nu$ and the two independent Gegenbauer function $C_{\nu-\mu}^\mu$ and $D_{\nu-\mu}^\mu$, the symmetries (\cite{DFS}, Sec.~3)
\ba
\label{C_symmetry}
C_{-\nu-\mu}^\mu(z) &=& -\frac{\sin{\pi(\nu+\mu)}}{\sin{\pi(\nu-\mu)}}C_{\nu-\mu}^\mu(z), \\
\label{D_symmetry}
    e^{i\pi(\mu-1)}D_{\nu+\mu-1}^{-\mu+1}(z) &=& e^{-i\pi\mu}\,2^{2\mu-1}\left(z^2-1\right)^{\mu-\frac{1}{2}}\frac{\Gamma(\nu-\mu+1)\Gamma(\mu)}{\Gamma(\nu+\mu)\Gamma(-\mu+1)}D_{\nu-\mu}^\mu(z), 
\ea
reduce the number of independent solutions for $w$ to four.  For $H^d$ we can take these as
\ba
\label{C_solutions}
\left(\sinh{\theta}\right)^l C_{i\sigma-l-\alpha}^{l+\alpha}(\cosh{\theta}),\quad && \left(\sinh{\theta}\right)^{-l-2\alpha+1} C_{i\sigma+l+\alpha-1}^{-l-\alpha+1}(\cosh{\theta}), \\
\label{D_solutions}
\left(\sinh{\theta}\right)^l D_{i\sigma-l-\alpha}^{l+\alpha}(\cosh{\theta}), \quad && \left(\sinh{\theta}\right)^{l} D_{-i\sigma-l-\alpha}^{l+\alpha}(\cosh{\theta}),
\ea
where $\sigma=\sqrt{k^2R^2-\alpha^2}$. The solutions for $S^d$ have the same form but  with the hyperbolic functions replaced by spherical functions and $i\sigma$ replaced by $\tau=\sqrt{k^2R^2+\alpha^2}$. 

The signs of $\alpha+l$ are different in the two  functions of the first kind in \eq{C_solutions}; the signs of $\sigma$ differ for the functions of the second kind in \eq{D_solutions}. The functions within the two sets are not connected by the symmetries in Eqs.~(\ref{C_symmetry}) and (\ref{D_symmetry}) which may only be used to change the signs of $\sigma$ and $\alpha+l$, respectively. We note also that possible solutions with the arguments replaced by their negatives are related to those in Eqs.~(\ref{C_solutions}) and (\ref{D_solutions}) by the reflection symettries of the Gegenbauer functions (\cite{DFS}, Sec.~5).

The solutions above are related to those in terms of associated Legendre functions used in \cite{CohlDangDunster} by the relations
\ba
\label{CPrelation}
C_{\nu-\mu}^\mu(z) &=& \sqrt{\pi}\,2^{-\mu+\frac{1}{2}}\frac{\Gamma(\nu+\mu)}{\Gamma(\mu)\Gamma(\nu-\mu+1)}\left(z^2-1\right)^{-\frac{\mu}{2}+\frac{1}{4}}P^{-\mu+\frac{1}{2}}_{\nu-\frac{1}{2}}(z), \\
\label{DQrelation}
D_{\nu-\mu}^\mu(z)  &=& \frac{1}{\sqrt{\pi}}e^{2\pi i(\mu-\frac{1}{4})}\,2^{-\mu+\frac{1}{2}}\frac{\Gamma(\nu+\mu)}{\Gamma(\mu)\Gamma(\nu-\mu+1)}\left(z^2-1\right)^{-\frac{\mu}{2}+\frac{1}{4}}Q^{-\mu+\frac{1}{2}}_{\nu-\frac{1}{2}}(z).
\ea
The symmetries of the Legendre functions ({\cite{dlmf}, Sec.~14.9(iii)), again limit the number of independent solutions to four.

Because of the dependence of the scalar Green functions on the composite angle $\Theta$ as in Eqs.~(\ref{ThetaH}) and (\ref{ThetaS}),  it will useful to note the addition formulas 
\ba
C_{\nu-\mu}^\mu(\cosh{\Theta}) &=& \frac{\Gamma(2\mu-1)}{\left[\Gamma(\mu)\right]^2}\sum_{n=0}^\infty(-1)^n\frac{2^{2n}\Gamma(\nu-\mu-n+1)\left[\Gamma(\mu+n)\right]^2}{\Gamma(\nu+\mu+n)} \nonumber \\
\label{Caddition}
&& \times(2n+2\mu-1)\left(\sinh{\theta}\sinh{\theta'}\right)^nC_{\nu-\mu-n}^{\mu+n}(\cosh{\theta})C_{\nu-\mu-n}^{\mu+n}(\cosh{\theta'})C_n^{\mu-\frac{1}{2}}(\cos{\varphi}), \\
D_{\nu-\mu}^\mu(\cosh{\Theta}) &=& \frac{\Gamma(2\mu-1)}{\left[\Gamma(\mu)\right]^2}\sum_{n=0}^\infty(-1)^n\frac{2^{2n}\Gamma(\nu-\mu-n+1)\left[\Gamma(\mu+n)\right]^2}{\Gamma(\nu+\mu+n)} \nonumber \\
\label{Daddition}
&& \times(2n+2\mu-1)\left(\sinh{\theta}\sinh{\theta'}\right)^nD_{\nu-\mu-n}^{\mu+n}(\cosh{\theta_>})C_{\nu-\mu-n}^{\mu+n}(\cosh{\theta_<})C_n^{\mu-\frac{1}{2}}(\cos{\varphi}),
\ea
where $\theta_>,\,\theta_<$ are the greater and lesser of $\theta$ and $\theta'$. These addition formulas and their ranges of validity are discussed in detail in \cite{DFS}, Sec. 8. See also  \cite{Vilenkin}, Chap.\ X, Sec.\ 3.5 and \cite{Henrici}. Integration over $\varphi$ using the orthogonality relations for the Gegenbauer polynomials (\cite{HTF}, 3.15.1 (16)-(20))  gives
\ba
\label{Cproduct}
\int_0^\pi C_{\nu-\mu}^\mu(\cosh{\Theta})(\sin{\varphi})^{2\mu-1}d\varphi &=& 2^{2\mu-1}\frac{\Gamma(\nu-\mu+1)\left[\Gamma(\mu)\right]^2}{\Gamma(\nu+\mu)}C_{\nu-\mu}^\mu(\cosh{\theta})C_{\nu-\mu}^\mu(\cosh{\theta'}), \\
\label{Dproduct}
\int_0^\pi D_{\nu-\mu}^\mu(\cosh{\Theta})(\sin{\varphi})^{2\mu-1}d\varphi &=& 2^{2\mu-1}\frac{\Gamma(\nu-\mu+1)\left[\Gamma(\mu)\right]^2}{\Gamma(\nu+\mu)}D_{\nu-\mu}^\mu(\cosh{\theta_>})C_{\nu-\mu}^\mu(\cosh{\theta_<}),
\ea
for the functions on $H^d$. The corresponding results for $S^d$ involve the replacement of hyperbolic functions and angles by spherical functions and angles throughout these expressions. The products of Gegenbauer functions on the right-hand sides of Eqs.~(\ref{Cproduct}) and(\ref{Dproduct}) will appear in the Mehler-Fock transforms introduced in Sec.~\ref{subsec:MehlerFock}.


\subsection{Conditions for the construction of the Green function \label{subsec:constructG}}

As was discussed at the beginning of Sec.~\ref{subsec:coordinates}, the only angular dependence of the scalar Green function is through the scalar product $x\cdot x'$ and the angle $\Theta$, ${\cal G}={\cal G}(\Theta,\omega)$. It can have no overall multiplicative dependence on angles through the hyperspherical harmonics $Y_l^M(\theta_1,\cdots,\theta_{d-1},\phi)$, so the angular momentum must vanish giving $l=0$.  We will assume the choice of coordinates given just before Eqs.~(\ref{ThetaH}) and (\ref{ThetaS}) and the expressions for $\Theta$ given in those equations.  To establish our methods, we will begin with the case of $H^d$. We will consider the case of $S^d$, which involves some further subtleties, in Sec.~\ref{sec:generalizedMehlerFock}.

 For $l=0$ and our choice of coordinates, there is no dependence on any angle except the angle $\varphi$ in \eq{ThetaH}.  To reduce the dependence of ${\cal G}(\Theta,\omega)$ from $\Theta$ to the radial variables $\theta$ and $\theta'$ alone, we will integrate over $\varphi$ and define the radial Green function $G(\theta,\theta',\omega)$ as
\be
\label{Gintegrated}
G(\theta,\theta',\omega)=\int_0^\pi {\cal G}(\Theta,\omega) d\varphi.
\ee
This function must satisfy the the inhomogeneous version of the radial wave equation, giving the defining relations
\ba
\label{Gequation}
&& \left[-\bigtriangleup_\theta-k^2R^2\right]G(\theta,\theta',\omega) = \delta(\theta-\theta')/(\sinh{\theta'})^{d-1}, \\
\label{Gdefined}
&& G(\theta,\theta',\omega) = \left[-\bigtriangleup_\theta-k^2R^2\right]^{-1} \delta(\theta-\theta')/(\sinh{\theta'})^{d-1}.
\ea
Here  $\bigtriangleup_\theta$ is the reduction of $R^2\bigtriangleup$ to the radial variable $\theta$ with $x_0=R\cosh{\theta}$, given for  $H^d$  by the negative of the expression in large  round parentheses in \eq{wave_eq_H}. The factor $1/(\sinh{\theta'})^{d-1}$ cancels the standard weight in integrations on $\theta'$. The full Green function including the $R$ dependence includes an overall factor $R^{-d}$ necessary to cancel the corresponding factor $R^d$ in the volume element in $d+1$ dimensions, a further factor $R^2$ connecting $\bigtriangleup_\theta$ to $\bigtriangleup$, and an angular normalization $1/\Omega$ where $\Omega$ is the total solid angle on $H^d$, $\Omega=2\pi^{d/2}/\Gamma(d/2)$. For simplicity we will suppress these factors until the end of Sec.~\ref{subsec:G_R}.

To proceed, we will  write the generalized Dirac delta distribution $\delta(\theta-\theta')/(\sinh{\theta'})^{d-1}$ in terms of the kernel of a Mehler-Fock-Gegenbauer transform of order $\alpha=(d-1)/2$ and its inverse. These transforms are a special case  of the Fourier-Jacobi transforms of Flensted-Jensen \cite{Flensted-Jensen}, Flensted-Jensen and Koornwinder \cite{FJ-Koornwinder}, and Koornwinder \cite{Koornwinder2}. With the delta distribution expressed in that form, the inverse operation in \eq{Gdefined} is simple to implement, and we can use complex integration both to obtain $G(\theta,\theta',\omega)$ and to impose the causality condition to obtain the retarded radial Green functions $G_R(\theta,\theta',\omega)$. We then invert the $\varphi$ integration to obtain ${\cal G}_R(\Theta,\omega)$.


\section{The Mehler-Fock transform and retarded Green functions on $H^d$ \label{sec:MehlerFock}}


\subsection{The generalized Mehler-Fock transform \label{subsec:MehlerFock}}


The generalized Mehler-Fock transform (\cite{dlmf}, Sec.\ 14.20(vi)) of a symmetric function $f(\theta)$ can be written in terms of Gegenbauer functions as
\be
\label{Mehler_trans}
\hat{f}(\lambda) = \int_0^\infty \frac{C_{i\lambda-\alpha}^\alpha(\cosh{\theta'})}{\sin[\pi(i\lambda-\alpha)]}f(\theta')\left(\sinh{\theta'}\right)^{2\alpha}d\theta',\quad 0\leq \theta<\infty,\quad \Re\alpha>-\frac{1}{2},
\ee
where the factor $(\sinh{\theta'})^{2\alpha}$ is the standard integration weight for the Gegenbauer functions of order $\alpha$.
The function $C_{i\lambda-\alpha}^\alpha(\cosh{\theta})/\sin[\pi(i\lambda-\alpha)]$ is a symmetric function of $\lambda$, so $\hat{f}$ is as well, $\hat{f}(-\lambda)=\hat{f}(\lambda)$.
The inverse transform is given by
\ba
\label{Mehler_inv}
f(\theta) &=& \int_0^\infty\hat{f}(\lambda) \frac{C_{i\lambda-\alpha}^\alpha(\cosh{\theta})}{\sin[\pi(i\lambda-\alpha)]} r(\lambda,\alpha) d\lambda, \\
\label{r(lambda,alpha)}
r(\lambda,\alpha) &=& 2^{2\alpha-1}\frac{\lambda\sinh{\pi\lambda}\left[\Gamma(\alpha)\right]^2}{\Gamma(-i\lambda+\alpha)\Gamma(i\lambda+\alpha)}.
\ea
This transform is a special case of the more general Fourier-Jacobi transform studied by  Flensted-Jensen  and Koornwinder \cite{Flensted-Jensen,FJ-Koornwinder,Koornwinder2}; see also (\cite{dlmf}, 15.9(ii)).  

We will express the function $C_{i\lambda-\alpha}^\alpha(\cosh{\theta})$ in \eq{Mehler_inv} in terms of Gegenbauer functions of the second kind,
\be
\label{CfromD}
\frac{C_{i\lambda-\alpha}^\alpha(\cosh{\theta})}{\sin[\pi(i\lambda-\alpha)]} = e^{-i\pi\alpha}\frac{1}{i\sinh{\pi\lambda}}\left[D_{i\lambda-\alpha}^\alpha(\cosh{\theta})-D_{-i\lambda-\alpha}^\alpha(\cosh{\theta})\right].
\ee
Then, using the symmetry of $\hat{f}(\lambda)$ and the reflection formula $\Gamma(z)\Gamma(1-z)=\pi/\sin{\pi z}$  for the gamma function (\cite{dlmf} Eq.~5.5.3), we can rewrite \eq{Mehler_inv} as \cite{LDaddition_formulas}
\be
\label{Mehler_inv_D}
f(\theta) = -ie^{-i\pi\alpha} \int_{-\infty}^\infty \hat{f}(\lambda)D_{i\lambda-\alpha}^\alpha(\cosh{\theta})r(\lambda,\alpha)\left[\sinh{\pi\lambda}\right]^{-1}d\lambda.
\ee
Equations (\ref{Mehler_trans}) and (\ref{Mehler_inv_D}) give a generalization of the Mehler-Fock transform as noted in  \cite{LDaddition_formulas}, Sec.\ 2.1.

Substituting the expression for $\hat{f}(\lambda)$ in \eq{Mehler_trans} into \eq{Mehler_inv_D}, we find that
\ba
f(\theta) &=&
\int_{0}^\infty \left[\frac{e^{-i\pi\alpha}}{2\pi}\int_{-\infty}^\infty D_{i\lambda-\alpha}^\alpha(\cosh{\theta})C_{i\lambda-\alpha}^\alpha(\cosh{\theta'}) \right. \nonumber \\ 
\label{Mehler_inv_D2}
&& \left.\times\frac{ 2^{2\alpha}i\lambda\left[\Gamma(\alpha)\right]^2\Gamma(i\lambda-\alpha+1)}{\Gamma(i\lambda+\alpha)} d\lambda\right]f(\theta')(\sinh{\theta'})^{2\alpha}d\theta'.
\ea
Given the reproducing property of this integral, the factor in square brackets clearly gives a representation of the generalized Dirac delta distribution $\delta(\theta-\theta')/(\sinh{\theta'})^{2\alpha}$, the expected radial delta distribution on $H^d$, (\cite{CohlDangDunster}, Sec.~4.2).

Importantly for the later construction of the retarded Green function, we can also divide the integration range in $\theta'$ in \eq{Mehler_inv_D2} into the ranges $0\leq\theta'<\theta$ and $\theta<\theta'<\infty$ and treat these separately in the $\lambda$ integration to obtain the expression we will use in Sec.~\ref{subsec:G_R},
\ba
f(\theta) &=&
\int_{0}^\infty \left[\frac{e^{-i\pi\alpha}}{\pi}\int_{-\infty}^\infty D_{i\lambda-\alpha}^\alpha(\cosh{\theta_>})C_{i\lambda-\alpha}^\alpha(\cosh{\theta_<}) \right. \nonumber \\ 
\label{Mehler_inv_D3}
&& \left.\times\frac{ 2^{2\alpha-1}i\lambda\left[\Gamma(\alpha)\right]^2\Gamma(i\lambda-\alpha+1)}{\Gamma(i\lambda+\alpha)} d\lambda\right]f(\theta')(\sinh{\theta'})^{2\alpha}d\theta',
\ea
where $\theta_>$ ($\theta_<)$ is the greater (lesser) of $\theta,\,\theta'$. The product of Gegenbauer functions in this expression is just that in \eq{Dproduct}. As a result, we could rewrite the integral in \eq{Mehler_inv_D3} as a double integral over $\lambda$ and the angle $\varphi$ in the composite angle $\Theta$ discussed before \eq{ThetaH}, with the product of Gegenbauer functions in the integrand replaced by $D_
{i\lambda-\alpha}^\alpha(\cosh{\Theta})$. Although we will not do this, the corresponding structure will be used in our analysis of the Green function in Sec.~\ref{subsec:G_R}.


\subsection{Properties of the kernel of the Mehler-Fock transform\label{subsec:MehlerFork_kernel_Hd}}


We will concentrate now on the properties of the kernel of the Mehler-Fock transform in \eq{Mehler_inv_D3},
 \ba
 && \frac{e^{-i\pi\alpha}}{2\pi}\int_{-\infty}^\infty D_{i\lambda-\alpha}^\alpha(\cosh{\theta_>})C_{i\lambda-\alpha}^\alpha(\cosh{\theta_<})  
\frac{ 2^{2\alpha}i\lambda\left[\Gamma(\alpha)\right]^2\Gamma(i\lambda-\alpha+1)}{\Gamma(i\lambda+\alpha)} d\lambda \nonumber \\
\label{Mehler_kern_Hd}
&& \quad =\delta(\theta-\theta')/(\sinh{\theta'})^{2\alpha}.
\ea
where $0<\theta,\,\theta'<\infty$. This relation is actually symmetric in $\theta$ and $\theta'$ as is evident from the first line.

To see the emergence of the delta distribution in \eq{Mehler_kern_Hd} explicitly, we will  use the asymptotic behavior of the Gegenbauer functions as functions of $\lambda$ to estimate the integral. This asymptotic behavior follows  from Watson's results on more general hypergeometric functions in \cite{HTF}, Sec.\ 2.3.2 (17), and was derived directly in \cite{DFS}, Sec.\ 6, and \cite{LDaddition_formulas}, Appendix, and in more detail in \cite{LDasymptotics}. It can also be extracted from the uniform asymptotic expansions for the associated Legendre functions derived in \cite{CohlDangDunster}, Sec.~2.3, using the connections in Eqs.~(\ref{CPrelation}) and (\ref{DQrelation}). Then with $z\in\mathbb{C}$ with $-\pi\leq \arg(z\pm 1)\leq\pi$,  $z_\pm=z\pm\sqrt{z^2-1}$, $z_-=1/z_+$,  $-\pi/2\leq\arg\nu\leq\pi/2$, $\Re\mu>0$, and $\lvert\nu\rvert\rightarrow\infty$,
\ba
\label{Dasymptotics}
D_{\nu-\mu}^\mu(z)&=&e^{i\pi\mu}\frac{2^{-\mu}}{\Gamma(\mu)}\nu^{\mu-1}\left(z^2-1\right)^{-\mu/2}z_+^{-\nu}\left[1+O(1/\lvert\nu\rvert)\right], \\
\label{Casymptotics}
C_{\nu-\mu}^\mu(z)&=&\frac{2^{-\mu}}{\Gamma(\mu)}\nu^{\mu-1}\left(z^2-1\right)^{-\mu/2}\left(e^{\pm i\pi\mu}z_+^{-\nu}+z_+^{\nu}\right)\left[1+O(1/\lvert\nu\rvert)\right],\quad \Im z\gtrless 0.
\ea
The asymptotic expression for $C_\nu^\mu(z)$ must be treated with care. One of the terms in \eq{Casymptotics} is often exponentially small relative to the other and to the error estimate, and should be dropped. For example, for $z$ real, $z\in(1,\infty)$, and $\Re\nu\gg 1$, the first term should be dropped; $C_\nu^\mu(z)$ then properly has no discontinuity across the real $z$ axis for $z>1$.

The results in Eqs.\ (\ref{Dasymptotics}) and (\ref{Casymptotics}) hold for $\lvert\nu\rvert\gg\lvert 2\sqrt{\mu}/\sqrt{z^2-1}\rvert$, so cannot be used for $z\rightarrow 1$ for fixed large $\lvert\nu\rvert$. That case is covered by  alternative asymptotic expansions of the Gegenbauer functions in terms of Bessel functions \cite{LDasymptotics,LD_Bessel_expansions} which give the correct limiting results for $z\rightarrow 1$, 
\ba
\label{Dasymptotics2}
D_{\nu-\mu}^\mu(z)&=&e^{i\pi\mu}\frac{1}{\sqrt{\pi}\,\Gamma(\mu)}2^{-\mu+\frac{1}{2}}\nu^{\mu-\frac{1}{2}}\left(z^2-1\right)^{-\frac{1}{2}\mu+\frac{1}{4}}K_{\mu-\frac{1}{2}}\left(Z\right)\left[1+O(1/\lvert\nu\rvert^{2/3}\right], \\
\label{Casymptotics2}
C_{\nu-\mu}^\mu(z) &=& \frac{\sqrt{\pi}}{\Gamma(\mu)}2^{-\mu+\frac{1}{2}}\nu^{\mu-\frac{1}{2}}\left(z^2-1\right)^{-\frac{1}{2}\mu+\frac{1}{4}} I_{\mu-\frac{1}{2}}(Z)\left[1+O(1/\lvert\nu\rvert^{2/3}\right],
\ea
where  $Z=\sqrt{2\nu^2(z-1)}$ and $\lvert\sqrt{z-1}\rvert\ll 1/\lvert\nu\rvert^{1/3}$.   The results in Eqs.\ (\ref{Dasymptotics}) and (\ref{Dasymptotics2}), and in (\ref{Casymptotics}) and (\ref{Casymptotics2}), agree in their common range of validity, $1/\lvert\nu\rvert\ll\sqrt{z-1}\ll\lvert\nu\rvert^{1/3}$, and agree also with the uniform asymptotic estimates in terms of Bessel functions derived by Cohl, Dang, and Dunster (\cite{CohlDangDunster}, Sec.\  2).

The asymptotic relations for  $C_{\nu+\mu}^{-\mu}(z)$ for $\lvert\nu\rvert\rightarrow\infty$ with $\Re\nu>0$ and $\Re\mu>0$ were not considered in \cite{LDasymptotics}. However, they can be extracted from the uniform asymptotic estimates for associated Legendre functions   in \cite{CohlDangDunster} using the relations in Eqs.~(\ref{CPrelation}) and (\ref{DQrelation}), and reduce to the result in \eq{Casymptotics} with $\mu\rightarrow-\mu$. 

We will now take $z=\cosh{\theta}\in(1,\infty)$, $z_+=e^{\theta}$, and use the asymptotic results in Eqs.\ (\ref{Dasymptotics}) and (\ref{Casymptotics}) in the expression in square brackets in \eq{Mehler_inv_D3} to estimate the integral. For  $\theta,\,\theta'\gg 1/\lvert\lambda\rvert$ with $\theta>\theta'$, this gives
\ba
\label{Mehler_kern_Hd2}
&& \frac{e^{-i\pi\alpha}}{2\pi}\int_{-\infty}^\infty  d\lambda\, D_{i\lambda-\alpha}^\alpha(\cosh{\theta})C_{i\lambda-\alpha}^\alpha(\cosh{\theta'})
 2^{2\alpha}i\lambda\left[\Gamma(\alpha)\right]^2\frac{\Gamma(i\lambda-\alpha+1)}{\Gamma(i\lambda+\alpha)} \\
\label{Mehler_kern_Hd3}
&&\quad \approx \left(\sinh{\theta'}\sinh{\theta}\right)^{-\alpha}\frac{1}{2\pi}\int_{-\infty}^\infty d\lambda\, e^{-i\lambda\theta}\left(e^{i\pi\alpha}e^{-i\lambda\theta'}+e^{i\lambda\theta'}\right) \\
\label{delta_function}
&& \quad= \delta(\theta-\theta')/(\sinh{\theta'})^{2\alpha} ,
\ea
where we have used a standard representation of the Dirac delta distribution,
\be
\label{delta_func}
\delta(x-x') = \frac{1}{2\pi}\int_{-\infty}^\infty d\lambda\, e^{i\lambda(x-x')},
\ee
 and recognized that the contributions to the exact integrals from the region near $\lambda=0$ are finite and do not affect the result \cite{footnote1}.
The corresponding calculation for $\theta<\theta'$ gives an identical result; $\delta(\theta-\theta')$ is even.
 
 For $\theta'$ and  $\theta$  both small, we cannot use the asymptotic approximations in Eqs.\ (\ref{Dasymptotics}) and (\ref{Casymptotics}), but must rather use the results in Eqs. (\ref{Dasymptotics2}) and(\ref{Casymptotics2}) or the corresponding uniform asymptotic expressions in \cite{CohlDangDunster}, Sec.\ 2.3. An estimate of the integral using Hankel's expansions for $K_\nu(z)$ and $I_\nu(z)$ for large arguments (\cite{dlmf}, Sec.\ 10.40) again reproduces the expected delta distribution, $\delta(\theta-\theta')/(\sinh{\theta'})^{2\alpha}$.

 As is evident from these calculations, the generalized Mehler-Fock  transforms in Eqs.~(\ref{Mehler_trans}) and (\ref{Mehler_inv}), or in Eqs.~(\ref{Mehler_trans}) and (\ref{Mehler_inv_D}), hold whether or not $\alpha$ is integer or half-integer as required for unitary representations of $H(d,1)/SO(d)$ on $H^d$. The order $\alpha$ of the Gegenbauer functions is  restricted only by the condition $\Re\alpha>-\frac{1}{2}$ for the validity of the original transform, \eq{Mehler_trans}.

We note for completeness that the Mehler-Fock kernel in \eq{Mehler_kern_Hd} can also be written as
\ba
&& -\frac{1}{2\pi}\int_{-\infty}^\infty d\lambda\, C_{i\lambda-\alpha}^\alpha(\cosh{\theta})C_{i\lambda-\alpha}^\alpha(\cosh{\theta'})\frac{ 2^{2\alpha-1}\lambda\sinh{\pi\lambda}\left[\Gamma(\alpha)\right]^2}{\sin{\left[\pi(i\lambda-\alpha)\right]}}\frac{\Gamma(i\lambda-\alpha+1)}{\Gamma(i\lambda+\alpha)} \nonumber \\ 
\label{alt_delta_func}
&&=\delta(\theta-\theta')/ \left(\sinh{\theta'}\right)^{2\alpha}.
\ea
These results follow  rigorously from the original form of the Mehler-Fock transform. As we will see, the form of the  kernel in \eq{alt_delta_func} cannot be used to construct retarded Green functions, but may be useful in other settings.


\subsection{Construction of the  retarded Green function on $H^d$ \label{subsec:G_R}}


The  frequency-dependent radial Green function for $H^d$ is given formally by the expression in \eq{Gdefined}, $G(\theta,\theta',\omega)=\left[-\bigtriangleup_\theta-k^2R^2\right]^{-1}\delta(\theta-\theta')/(\sinh{\theta'})^{2\alpha}$, subject to the causality or retardation condition.  Using the expression for the delta distribution in \eq{delta_func} and evaluating of the action of the  inverse operator on $D_{i\lambda-\alpha}^\alpha(\cosh{\theta})$ we find that 
\ba
G(\theta,\theta',\omega) &=& \frac{e^{-i\pi\alpha}}{2\pi}\int_{-\infty}^\infty  d\lambda\frac{D_{i\lambda-\alpha}^\alpha(\cosh{\theta_>})C_{i\lambda-\alpha}^\alpha(\cosh{\theta_<})}{-(i\lambda)^2-k^2R^2+\alpha^2}  \nonumber \\
\label{GH1}
&& \times\, 2^{2\alpha}i\lambda\left[\Gamma(\alpha)\right]^2\frac{\Gamma(i\lambda-\alpha+1)}{\Gamma(i\lambda+\alpha)}. 
\ea
The integrand has simple poles at $\lambda=\pm \sqrt{k^2R^2-\alpha^2}$ from the denominator, and at $\lambda=i(\alpha+n)$, $n=0,\,1,\,\cdots$ from the combination of the factors in \eq{GH1} with the coefficients of the hypergeometric functions in the definitions of the Gegenbauer functions,  Eqs.~(\ref{Cdefined}) and (\ref{Ddefined}). The original contour in $\lambda$ in \eq{Mehler_kern_Hd} can be distorted to run from $-\infty$ to $\infty$ in a finite strip in $\Im\lambda$ with $\Re\lambda<\alpha$, so can be taken to run above, below, or between the poles of the denominator. The proper contour will be determined by the retardation condition. 

The integrand in \eq{GH1} behaves asymptotically for $\lvert\lambda\rvert\rightarrow\infty$ as $e^{-i\lambda(\theta_>-\theta_<)}/\lambda^2$ so vanishes exponentially as a function of $\lambda$ for $\Im\lambda\rightarrow-\infty$. We can therefore close the $\lambda$ integration contour in \eq{GH1} in the lower half $\lambda$ plane. The result vanishes identically for an initial contour below the poles  and otherwise  can be expressed in terms of the residues at the poles. With  the definition $\sigma\equiv\sqrt{k^2R^2-\alpha^2}$, the poles at $\lambda=\pm\sigma$ give
\ba
G_\pm(\theta,\theta',\omega) &=& e^{-i\pi\alpha}2^{2\alpha-1}\left[\Gamma(\alpha)\right]^2\frac{\Gamma\left(\pm i\sigma-\alpha+1\right)}{\Gamma\left(\pm i\sigma+\alpha\right)} \nonumber \\
\label{GH2}
&& \times D_{\pm i\sigma-\alpha}^\alpha(\cosh{\theta_>})C_{\pm i\sigma-\alpha}^\alpha(\cosh{\theta_<}).
\ea

The time-dependent form of the radial Green function is given by the inverse Fourier transform in \eq{f_t},
\be
\label{GH3}
{\mathfrak G}_\pm(\theta,\theta',t) = \frac{1}{2\pi}\int_{-\infty}^\infty d\omega\, G_\pm(\theta,\theta',\omega) e^{-i\omega t},
\ee
where the contour in the $\omega$ integration must be chosen to provide a causal Green function. In particular, for a source on $H^d$ at $\theta'$, the integral in \eq{GH3} must vanish at points  geodesic distances $R(\theta_>-\theta_<)=R\lvert\theta-\theta'\rvert$ greater than $ct$ from the source. We will consider this separate $\omega$ integration in detail.

In the case of $G_+(\theta,\theta',\omega)$, the Gegenbauer functions in \eq{GH2} have poles in the upper half $\omega$ plane at  $\sigma=i(\alpha+n)$, $n=0,\,1, \ldots$, while for $\lvert\omega\rvert\rightarrow\infty$ in the lower half plane, $\lvert kR\rvert\rightarrow\infty$, and the integrand in \eq{GH3} behaves  $(1/\omega)e^{-i \omega/c)[ R(\theta_>-\theta'_<)+ct]}$ as a function of $\omega$. This function vanishes exponentially for $\Im\omega\rightarrow-\infty$  for $t>0$. We can therefore close the contour in the lower half $\omega$ plane, and find that ${\mathfrak G}_+$ vanishes in a region that includes the physical region $0<R(\theta_>-\theta_<)<ct$ for a retarded Green function ${\mathfrak G}_R$. ${\mathfrak G}_+$ therefore cannot contribute to ${\mathfrak G}_R$. We note that we could not close the contour and the result would be nonzero for $R(\theta_>-\theta_<)+ct<0$, corresponding to an advanced rather than retarded condition on ${\mathfrak G}_+$.

The contribution of $G_-(\theta,\theta',\omega)$ to the retarded Green function does not vanish. In this case, the poles of the Gegenbauer functions in the $\omega$ integration are in the lower half $\omega$ plane at  $\sigma=-i(\alpha+n)$, $n=0,\,1,\ldots$. The integrand behaves asymptotically as $(1/\omega)e^{i(\omega/c)[R(\theta_>-\theta_<)-\omega t]}$ for $\lvert\omega\rvert\rightarrow\infty$ in the upper half $\omega$ plane, so decreases exponentially for $\Im\omega\rightarrow+\infty$. We can therefore close the integration contour in the upper half plane, and find that  ${\mathfrak G}_-(\theta,\theta',t)$ vanishes for $R(\theta_>-\theta_<)>ct$. This is just the causality condition. In contrast, $\mathfrak{G}_-$ is non-zero for $R(\theta_>-\theta_<)-ct<0$. In that case, we can distort the contour to run around the singularities in the lower half $\omega$ plane. These are poles in $\sigma$, but singular branch points in $\omega$. The resulting integral over $\omega$  can apparently not be evaluated in closed form but does not vanish.  We conclude that the retarded Green function is ${\mathfrak G}_R={\mathfrak G}_-$.

This analysis shows that, to obtain the retarded Green function, the integration contour in $\lambda$ in \eq{GH1} must be chosen to run above the pole of the integrand at $\lambda=-\sqrt{k^2R^2-\alpha^2}$, but below the pole at $\lambda=+\sqrt{k^2R^2-\alpha^2}$, thus picking out only the contribution to $G_-(\theta,\theta',\omega)$ from the former when the contour of the $\lambda$ integration in \eq{GH1} is closed in the lower half $\lambda$ plane. Alternatively, we may take $\omega\rightarrow \omega-i\epsilon$, integrate on the real axis in $\lambda$, and let $\epsilon\rightarrow$ at the conclusion of the calculation.

Equation (\ref{alt_delta_func}) gives an alternative expression for the delta distribution $\delta(\theta-\theta')/(\sinh{\theta'})^{2\alpha}$ in which the angles $\theta$ and $\theta'$ appear symmetrically and the factor  $D_{i\lambda-\alpha}^\alpha(\cosh{\theta_>})$ in \eq{Mehler_kern_Hd} is replaced by $\frac{1}{2}C_{i\lambda-\alpha}^\alpha(\cosh{\theta})$. The two expressions are completely equivalent as far as the generalized Mehler-Fock transform is concerned. They are not equivalent for the construction of the Green functions: the asymptotic result for $C_{i\lambda-\alpha}^\alpha(\cosh{\theta})$ in \eq{Casymptotics} involves the simultaneous appearance of exponentials $e^{\pm i(\omega/c)R\theta}$ of both signs at each stage in the discussion of the $\omega$ integration above, and it is not possible to construct a retarded Green function using that form.

We conclude that the frequency-dependent retarded  Green function, integrated over $\varphi$, is
\ba
G_R(\theta,\theta',\omega) &=&  e^{-i\pi\alpha}2^{2\alpha-1}\left[\Gamma(\alpha)\right]^2\frac{\Gamma\left(-i\sigma-\alpha+1\right)}{\Gamma\left(- i\sigma+\alpha\right)} \nonumber \\
\label{GRH1}
&& \times D_{- i\sigma-\alpha}^\alpha(\cosh{\theta_>})C_{- i\sigma-\alpha}^\alpha(\cosh{\theta_<}), \quad \sigma=\sqrt{k^2R^2-\alpha^2}.
\ea

It will be useful for later purposes to show directly that this expression satisfies \eq{Gequation}. For this purpose we will rewrite the wave equation for $l=0$, \eq{wave_eq_H}, as 
\be
\label{wave_eq_H2}
\left(-\frac{1}{(\sinh{\theta})^{2\alpha}}\frac{d}{d\theta}(\sinh{\theta})^{2\alpha}\frac{d}{d\theta}-k^2R^2\right)w(\theta)=0.
\ee
The operator in this expression gives zero when acting on either of the Gegenbauer functions in \eq{GRH1} depending on whether $\theta\gtrless\theta'$ except at $\theta=\theta'$ where there is a discontinuity in the first derivative. In particular, the difference between the first derivative for $\theta>\theta'$ and that for $\theta<\theta'$ for $\theta\rightarrow\theta'$ is just the Wronskian for the Gegenbauer functions considered as functions of $\theta$ (\cite{dlmf} Eq.~14.2.10, and Eqs.~(\ref{CPrelation}) and (\ref{DQrelation})),
 \ba
 {\cal W}(D_{\nu-\mu}^\mu,C_{\nu-
 \mu}^\mu)_\theta &\equiv& D_{\nu-\mu}^\mu(\cosh{\theta})\frac{d}{d\theta}C_{\nu-\mu}^\mu(\cosh{\theta})-C_{\nu-\mu}^\mu(\cosh{\theta})\frac{d}{d\theta}D_{\nu-\mu}^\mu(\cosh{\theta} )\nonumber \\
 \label{W(D,C)}
 &=& e^{i\pi\mu}2^{2\mu-1}\frac{\Gamma(\nu+\mu)}{\left[\Gamma(\mu)\right]^2\Gamma(\nu-\mu+1)}(\sinh{\theta})^{-2\mu},
 \ea
with ${\cal W}$ evaluated for $\nu=-i\sigma$ and $\mu=\alpha$. With the additional factors in Eqs.~(\ref{GRH1}) and (\ref{wave_eq_H2}), we find a unit step function $\vartheta(\theta-\theta')$ at $\theta=\theta'$,  with $\vartheta(x)=1,\, x>0$ and $\vartheta(x)=0,\, x<0$.  The remaining derivative in \eq{wave_eq_H2} gives
 \be
 \label{delta_from_W}
 (\sinh{\theta})^{-2\alpha}\frac{d}{d\theta}\vartheta(\theta-\theta') = \delta(\theta-\theta')/(\sinh{\theta})^{2\alpha}
 \ee
 as expected from  \eq{Gequation} (see {\em e.g.} \cite {dlmf}, Eq.~(1.16.16)).

 
 \subsection{The scalar Green function \label{subsec:scalar_Greens_func}}
 

As we discussed in Sec.~\ref{subsec:constructG}, the scalar Green function ${\cal G}_R$ must be a function only of $\Theta$.  That condition together with the integral relation in \eq{Dproduct} and the result in \eq{GRH1} show that ${\cal G}_R(\Theta,\omega)$ is  proportional to $e^{-i\pi\alpha} D_{- i\sigma-\alpha}^\alpha(\cosh{\Theta})$ up to the possible addition of solutions of the homogeneous wave equation for $l=0$, \eq{wave_eq_H}, as functions of $\cosh{\Theta}$. 
Thus, taking $\alpha=(d-1)/2$, incorporating the factor $R^{2-d}$ discussed following \eq{Gdefined}, and dividing the result  by the total solid angle $\Omega=2\pi^{(d-1)/2}/\Gamma(\frac{1}{2}(d-1))=2\pi^\alpha/\Gamma(\alpha)$ on $H^d$ to account for the implied integration over the remaining angles that do not appear for $l=0$ and our choice of coordinates, we find that
\be
\label{GRH2}
{\cal G}_R^d(\Theta,\omega) = e^{-i\pi\alpha} \frac{\Gamma(\alpha)}{2\pi^{\alpha}R^{2\alpha-1}} D_{- i\sqrt{k^2R^2-\alpha^2}-\alpha}^{\alpha}(\cosh{\Theta})
\ee
with $\cosh{\Theta}=\cosh{\theta}\cosh{\theta'}-\sinh{\theta}\sinh{\theta'}\cos{\varphi}$.

This result for the Green function is unique. A possible choice of the four independent solutions of the homogeneous wave equation that could be added to this expression without changing the right hand side of \eq{Gdefined} is given in Eqs.~(\ref{C_solutions}) and (\ref{D_solutions}). The coefficient of $D_{i\sigma-\alpha}^\alpha$  is already fixed by \eq{GRH2}. The possible addition of $C_{i\sigma-\alpha}^\alpha$ is precluded by the retardation condition as discussed above. The same problem, the appearance of exponentials $e^{\pm i(\omega/c)R\Theta}$  with both signs in $\omega$, occurs for the second solution $C_{-i\lambda+\alpha-1}^{-\alpha+1}$ in \eq{C_solutions}, so its addition to  \eq{GRH2} is again precluded by causality. Finally, the second solution in \eq{D_solutions} leads in the $\omega$ integration to an exponential $e^{-i(kR\Theta+\omega ct)}$ and an advanced rather than retarded Green function. These functions can of course appear in the general solution to the wave equation in the presence of radiation not emitted by the source, just not in ${\cal G}_R^d(\Theta,\omega)$ itself \cite{outgoing_waveBC}.

The result in \eq{GRH2} is identical to that given by Cohl, Dang, and Dunster  in terms of associated Legendre functions (\cite{CohlDangDunster}, Theorem 4.6). This may be shown by using the relation in  \eq{DQrelation} and the symmetry of the functions $Q_\nu^\mu(z)$ for $\mu\rightarrow-\mu$  (\cite{dlmf}, Sec.~14.9(iii)). The expression in \eq{GRH2} therefore reduces properly for a source at $\theta'=0$ to the known results for the Green functions in the Euclidean spaces $E^d$ in the flat-space limit as shown by those authors. This limit corresponds physically to  high  enough frequencies or short enough wavelengths that $kR\Theta\gg1$ for $\Theta\ll1$. The solutions of the wave equation on $H^d$ and on its tangent space at $\Theta=0$ then do not differ significantly, with many wavelengths on either over a distance $R\Theta$ within which the geometries of the two spaces are essentially equivalent \cite{footnote2}.  

The physical interpretation of the product ${\cal G}_R(\Theta,\omega)e^{-i\omega t}$ is of some interest. For $kR\gg \alpha$ and $kR\Theta=k\lvert x-x'\rvert\gg 1$
\be
\label{Hpropagate}
{\cal G}_R^d(\Theta,\omega)e^{-i\omega t} \sim e^{i\frac{\pi}{2}(\alpha-1)}\frac{k^{\alpha-1}}{2^{\alpha+1}\pi^\alpha R^\alpha}(\sinh{\Theta})^{-\alpha}e^{ikR\Theta-i\omega t}                
\ee
and the product describes a wave propagating on $H^d$ at the speed of light. However, for $kR<\alpha$, the square root in \eq{GRH2} becomes imaginary, $\sqrt{k^2R^2-\alpha^2}\rightarrow i\sqrt{\alpha^2-k^2R^2}$, and the corresponding function
\be
\label{H_nonpropagate}
{\cal G}_R^d(\Theta,\omega)e^{-i\omega t}  \sim \frac{1}{2\pi^\alpha R^{2\alpha-1}}\frac{\Gamma (\sqrt{\alpha^2-k^2R^2})}{\Gamma(\sqrt{\alpha^2-k^2R^2}+1)}e^{-\sqrt{\alpha^2-k^2R^2}\Theta-i\omega t}    
\ee
describes a compact oscillating but non-propagating distribution at long wavelengths with the Gegenbauer function in \eq{GRH2} decaying exponentially for $\sqrt{\alpha^2-k^2R^2}\,\Theta\gg 1$ (\cite{DFS}, Eq.~2.2).

We emphasize that our approach has been quite different from that of \cite{CohlDangDunster}. We have derived the Green function directly for general $\alpha$ with $\Re\alpha>-\frac{1}{2}$. In particular, the result in \eq{GRH2} holds for non-integer $d$ with $\Re d>0$. In that case, there are no angles defined by the geometry. The angle $\varphi$ in $\cosh{\Theta}$ appears an auxiliary parameter used to connect Eqs. (\ref{GRH1}) and (\ref{GRH2}), while the total solid angle $\Omega=2\pi^{\alpha}/\Gamma(\alpha)$ and the factor $R^{d-2}=R^{2\alpha-1}$ divided out in \eq{GRH2} are the continuations from their values for integer $d$. This approach is common in the use of dimensional regularization in quantum field theory, and  ${\cal G}_ R^d(\Theta,\omega)$ for non-integer $d$ is in that sense the dimensional continuation  of the physical Green function for integer $d$. 

For $d$ integer and the choice of coordinates discussed before \eq{ThetaH}, $\cosh{\Theta}=x\cdot x'/R^2$. This scalar expression is unchanged by hyperbolic rotations, and
\be
\label{GRH3}
 {\cal G}_R^d(x,x',\omega) = e^{-i\pi(d-1)/2} \frac{\Gamma((d-1)/2)}{2R^{d-2}\pi^{(d-1)/2}} D_{- i\sqrt{k^2R^2-((d-1)/2)^2}-(d-1)/2}^{(d-1)/2}(x\cdot x'/R^2).
\ee
for arbitrary locations of $x,\,x'$ on $H^d$ with the separation $\lvert x-x'\rvert$ fixed. In an angular description, ${\cal G}_R^d$ then depends in general on all the angles $\theta_1,\cdots,\theta_{d-1},\,\phi$ in \eq{Hd_coord}. The full Laplacian including those angles then appears in the wave equation and the defining relation for $G$, and the delta distribution $\delta(\theta-\theta')/(\sinh{\theta'})^{d-1}$ in \eq{Gequation} must be generalized to include all angles as discussed in \cite{CohlDangDunster}, Sec.~4.2. The results in terms of $x,\,x'$ remain simple.


\section{ The retarded Green function and a new generalized Mehler-Fock transform on $S^d$ \label{sec:generalizedMehlerFock}}


\subsection{A generalized Mehler-Fock kernel for z\,=\,cos$\,\theta\in(-1,1)$\label{subsec:MehlerFock_Sd}}


To treat the case of $S^d$ where $0<\theta,\,\theta'<\pi$, we will begin by deriving an apparently new generalization of the Mehler-Fock kernel applicable to this case. Our method depends on our ability to continue the kernel distribution defined in \eq{Mehler_kern_Hd} for  $\cosh{\theta},\,\cosh{\theta'}\in(1,\infty)$ to the angular region of interest. We begin with the expression for the Mehler-Fock kernel in \eq{Mehler_kern_Hd} with $i\lambda$ replaced by a new variable $\nu$:
\be
\label{Mehler_kern}
\frac{\delta(\theta-\theta')}{(\sinh{\theta'})^{2\alpha}} = \frac{e^{-i\pi\alpha}}{2\pi i}\int_{-i\infty}^{i\infty}d\nu D_{\nu-\alpha}^\alpha(\cosh{\theta_>})C_{\nu-\alpha}^\alpha(\cosh{\theta_<})\frac{2^{2\alpha}\nu\left[\Gamma(\alpha)\right]^2\Gamma(\nu-\alpha+1)}{\Gamma(\nu+\alpha)}. 
\ee
The integrand in this expression has simple poles at $\nu=-\alpha,\,-\alpha-1,\cdots$ from the poles of the Gegenbauer functions, Eqs.~(\ref{Cdefined}) and (\ref{Ddefined}), and vanishes for $\Re\nu\rightarrow\infty$ in the right-half plane proportionally to $e^{-\nu(\theta_>\mp\theta_<)}/(\sinh{\theta}\sinh{\theta'})^{2\alpha}$ as seen from Eqs.~(\ref{Dasymptotics}) and (\ref{Casymptotics}).  The Gegenbauer function of the second kind is cut along the real axis for $z=\cosh{\theta_>}\leq 1$, with $z_+=z+\sqrt{z^2-1}\rightarrow e^{\pm i\theta}$ for $z\rightarrow x\pm i0$ with $x\in(-1,1)$. $C_{\nu-\alpha}^\alpha(z)$ is continuous across the interval $-1<x\leq 1$,  $C_{\nu-\alpha}^\alpha(x+i0)=C_{\nu-\alpha}^\alpha(x-i0)=C_{\nu-\alpha}^\alpha(\cos{\theta})$.
 
To transform  \eq{Mehler_kern} from $H^d$ to $S^d$, we note first that the wave equation on $S^d$, \eq{wave_eq_S},   follows from the Helmholtz equation on $H^d$ obtained by replacing $\omega$ in \eq{wave_eq_H} by  $\mp i\omega$,  by making a transformation of $\theta$ with the complementary phase, $\theta\rightarrow\pm i\theta$. We will therefore continue the expression  in \eq{Mehler_kern} simultaneously in $\theta,\,\theta'$ and $\nu$ keeping the phases of $\theta$ and $\theta'$ the same, and the changing the phase of $\nu$ in the opposite sense to preserve the asymptotic structure for $\lvert\nu\rvert\rightarrow\infty$ \cite{footnote3}. 
Thus for $z=\cosh{\theta}\rightarrow x\pm i0$ and $z'=\cosh{\theta'}\rightarrow x'\pm i0$, with $x=\cos{\theta}$ and $x'=\cos{\theta'}$, we continue $\nu$ as $\nu\rightarrow e^{\mp i\pi/2}\nu$. The left-hand side of \eq{Mehler_kern} then continues as
\be
\label{delta_cont}
\delta(\theta-\theta')/(\sinh{\theta'})^{2\alpha}\rightarrow \delta(\pm i(\theta-\theta'))/(e^{\pm i\pi/2}\sin{\theta'})^{2\alpha} = \mp ie^{\mp i\pi\alpha}\delta(\theta-\theta')/(\sin{\theta'})^{2\alpha}
\ee
using the standard relation $\delta(ax)=\delta(x)/a$. The continuation in $\nu$ leads to a rotation in the integration contour in \eq{Mehler_kern} by $\mp \pi/2$ giving the integrals
\be
\label{Mehler_kern2}
\pm\frac{e^{-i\pi\alpha}}{2\pi i}\int_{C_\pm}d\nu D_{\nu-\alpha}^\alpha(\cos{\theta_>}\pm i0)C_{\nu-\alpha}^\alpha(\cos{\theta_<})\frac{2^{2\alpha}\nu\left[\Gamma(\alpha)\right]^2\Gamma(\nu-\alpha+1)}{\Gamma(\nu+\alpha)},
\ee
where the contours $C_\pm$ run from $-\infty$ to $+\infty$ in finite strips passing below $(+)$ or above $(-)$ the negative real $\nu$ axis.

Combining Eqs.\ (\ref{delta_cont}) and (\ref{Mehler_kern2}) we obtain our basic expression for the Mehler-Fock kernel on $S^d$,
\be
\label{Mehler_kern3}
\delta(\theta-\theta')/(\sin{\theta'})^{2\alpha} =  e^{\pm i\pi\alpha}\frac{e^{-i\pi\alpha}}{2\pi}\int_{C_\pm}d\nu D_{\nu-\alpha}^\alpha(\cos{\theta_>}\pm i0)C_{\nu-\alpha}^\alpha(\cos{\theta_<})\frac{2^{2\alpha}\nu\left[\Gamma(\alpha)\right]^2\Gamma(\nu-\alpha+1)}{\Gamma(\nu+\alpha)}.
\ee
This expression is symmetric in $\theta$ and $\theta'$, and is valid for $0< \theta, \theta'<\pi$.

We can check this result approximately by using the asymptotic forms for the Gegenbauer functions for $\lvert\nu\rvert\rightarrow\infty$, Eqs.~(\ref{Dasymptotics}) and (\ref{Casymptotics}) and Thm.\ 2 in \cite{LDasymptotics}. Thus for $\theta,\,\theta'<\pi/2$ not too small,
\ba
\label{Dasymptotics3}
D_{\nu-\alpha}^\alpha(\cos{\theta}\pm i0) &\sim& e^{i\pi\alpha}e^{\mp i\pi\alpha}\frac{2^{-\alpha}}{\Gamma(\alpha)}\nu^{\alpha-1}(\sin{\theta})^{-\alpha}e^{\mp i(\nu\theta-\pi\alpha/2)}, \\
\label{Casymptotics3}
C_{\nu-\alpha}^\alpha(\cos{\theta}) &\sim& \frac{2^{-\alpha+1}}{\Gamma(\alpha)}\nu^{\alpha-1}(\sin{\theta})^{-\alpha}\cos{\left(\nu\theta-{\pi\alpha/2}\right)},
\ea
$\lvert\nu\rvert\rightarrow\infty$. These results and their extensions \cite{LDasymptotics} to other ranges of $\theta$ in the interval $0<\theta<\pi$  show that the integrands in \eq{Mehler_kern3} are analytic in the lower (upper) half $\nu$ planes for $\Im\cos{\theta},\Im\cos{\theta'} \gtrless 0$ and vanish for $\Im\nu\rightarrow -\infty$ ($+\infty$) in those half planes as expected from our construction. They show furthermore that the integral in \eq{Mehler_kern3} contains the expected delta distribution. Thus, for the conditions under which the asymptotic approximations in Eqs.~(\ref{Dasymptotics3}) and (\ref{Casymptotics3}) apply, the right-hand side of \eq{Mehler_kern3} gives
\be
\label{delta_on_Sd}
(\sin{\theta}\sin{\theta'})^{-\alpha}\int_{-\infty}^\infty d\nu\left( e^{(\mp i\nu- a)(\theta_>-\theta_<)}+e^{\pm i\pi\alpha}e^{(\mp i\nu-a)(\theta_>+\theta_<)}\right) = \delta(\theta-\theta')/(\sin{\theta'})^{2\alpha}
\ee
when integrated on contours $C_\pm$ a distance $a$ below $(+)$ or above $(-)$ the real axis, away from the poles on the negative real axis at $\nu=-\alpha,\,-\alpha-1,\cdots$. Note that the dependence on $a$ disappears in the first delta-function term as $\theta\rightarrow\theta'$, while the second $a$-dependent term vanishes as a distribution. We will present a direct derivation of the relation in \eq{Mehler_kern3} in Sec.~\ref{subsec:MehlerFock_onSd}.


\subsection{Construction of the retarded Green  function on $S^d$ \label{subsec:retardedGonSd}}


The frequency-dependent radial Green function for $S^d$ is given by the analog of the expression in \eq{Gdefined} with, in this case, $-\bigtriangleup_\theta w_{\nu-\alpha}^\alpha=(\nu^2-\alpha^2)w_{\nu-\alpha}^\alpha$. Here $w_{\nu-\alpha}^\alpha(\cos{\theta})$  again a Gegenbauer function of either the first or second kind. Thus,
\ba
G_\pm(\theta,\theta',\omega) &=& [-\bigtriangleup_\theta-k^2R^2]^{-1}\delta(\theta-\theta')/(\sin{\theta'})^{2\alpha} \\
&=& e^{\pm i\pi\alpha}\frac{e^{-i\pi\alpha}}{2\pi}\int_{C_\pm} d\nu\frac{D_{\nu-\alpha}^\alpha(\cos{\theta_>}\pm i0) C_{\nu-\alpha}^\alpha(\cos{\theta_<)}}{\nu^2-k^2R^2-\alpha^2} \nonumber \\
\label{GonSdefined}
&& \times \frac{2^{2\alpha}\nu[\Gamma(\alpha)]^2\Gamma(\nu-\alpha+1)}{\Gamma\nu+\alpha)},
\ea
$\alpha=(d-1)/2$. The integrand has simple poles in $\nu$ at $\nu=\pm\sqrt{k^2R^2+\alpha^2}$ from the zeros of the denominator, and at $\nu=-\alpha-n$, $n=0,\,1,\ldots,$ from factors in the Gegenbauer functions,  Eqs.~(\ref{Cdefined}) and (\ref{Ddefined}). The integration  contours $C_\pm$ again run from $-\infty$ to $+\infty$ passing  below $(+)$  or above $(-)$ the negative real axis for $\cos{\theta}\pm i0$. 

In the case of $C_-$ the integrand in \eq{GonSdefined} behaves asymptotically as $e^{i\nu(\theta_>-\theta_<)}/\nu^2$ for $\Im\nu\rightarrow\infty$, so the contour can be closed with a loop at infinity and integral vanishes identically unless at least one of the poles of the denominator is inside. We start with the pole at $\nu=\sqrt{k^2R^2+\alpha^2}$, supposing that this is displaced slightly into the upper half $\nu$ plane so that $C_-$ runs below it. The residue of the pole gives
\ba
G_-(\theta,\theta',\omega) &=& ie^{-2\pi i\alpha}2^{2\alpha-1}[\Gamma(\alpha)]^2\frac{\Gamma(\tau-\alpha+1)}{\Gamma(\tau+\alpha)}
\nonumber \\
\label{GonSdefined2}
&&\times D_{\tau-\alpha}^\alpha(\cos{\theta_>}-i0)C_{\tau-\alpha}^\alpha(\cos{\theta_<})
\ea
with $\tau=\sqrt{k^2R^2+\alpha^2}$.  This choice of the pole corresponds to the continuation for $\omega\rightarrow e^{ i\pi/2}\omega$  complementary to the angular continuations $\theta,\,\theta'\rightarrow e^{- i\pi/2}\theta,\, e^{- i\pi/2}\theta'$ of the pole at $i\lambda=-\sqrt{k^2R^2-\alpha^2}$ that gave the retarded Green function on $H^d$. 

The time-dependent form of the radial Green function is given by the inverse Fourier transformation in \eq{f_t} or \eq{GH3}. The function $G_-(\theta,\theta',\omega)$ behaves as $e^{(i\omega/c)[R(\theta_>-\theta_<)-ct]}/\omega$ for $\Im\omega\rightarrow\infty$. The contour can be closed in the upper half $\omega$ plane for $R(\theta_>-\theta_<)>ct$, the integral vanishes, and the retardation condition is satisfied. The integral does not vanish for $R(\theta_>-\theta_<)<ct$. Similar considerations show that the retardation condition is not satisfied  by the contributions of the pole at $\nu=-\sqrt{k^2R^2+\alpha^2}$ for integration on either $C_-$, or $C_+$, while the the pole at $\nu=\sqrt{k^2R^2+\alpha^2}$ gives an advanced rather than retarded contribution with respect to integration on $C_+$. 

We conclude that $G_-(\theta,\theta',\omega)$, \eq{GonSdefined2}, gives the frequency-dependent form of the retarded radial Green function $G_R(\theta,\theta',\omega)$. This result is unique. It is not possible to add solutions of the homogeneous wave equation without changing the normalization of $G_-(\theta,\theta',\omega)$ or violating the retardation condition.

 $G_R(\theta,\theta',\omega)$ is the integral over angles of the scalar Green function as is evident for our choice of coordinates from the analogs of the relations in Eqs.~(\ref{Daddition}) and (\ref{Dproduct}) with the hyperbolic angles replaced by spherical  angles. Thus
\ba
 && 2^{2\alpha-1}\frac{\Gamma(\nu-\alpha+1)\left[\Gamma(\alpha)\right]^2}{\Gamma(\nu+\alpha)}D_{\tau-\alpha}^\alpha(\cos{\theta_>}-i0)C_{\tau-\alpha}^\alpha(\cos{\theta_<}) \nonumber \\ 
 \label{Dproduct2}
 && = \int_0^\pi D_{\tau-\alpha}^\alpha(\cos{\Theta-i0})(\sin{\varphi})^{2\alpha-1}d\varphi
\ea
with $\cos{\Theta}=\cos{\theta}\cos{\theta'}+\sin{\theta}\sin{\theta'}\cos{\varphi}$. We obtain the scalar Green function on $S^d$ by dropping the integral over $\varphi$ and supplying the factors of $R$ and solid angle discussed preceding \eq{GRH2}, 
\be
\label{GRS1}
{\cal G}_R^d(\Theta,\omega) = i e^{-2\pi i\alpha} \frac{\Gamma(\alpha)}{2\pi^{\alpha}R^{2\alpha-1}} D_{\sqrt{k^2R^2+\alpha^2}-\alpha}^{\alpha}(\cos{\Theta}-i0).
\ee
Alternatively, in terms of the coordinates $x\,,x'$, 
\be
\label{GRS2}
{\cal G}_R^d(x,x',\omega) =i e^{-2\pi i\alpha} \frac{\Gamma(\alpha)}{2\pi^{\alpha}R^{2\alpha-1}} D_{\sqrt{k^2R^2+\alpha^2}-\alpha}^{\alpha}\left(\frac{x\cdot x'}{R^2}-i0\right).
\ee

The limit of this expression for $kR\gg\alpha$ and $\Theta<\pi$,
\be
\label{GRSprop}
{\cal G}_R^d(\Theta,\omega)e^{-i\omega t} \sim ie^{-i\pi\alpha/2}\frac{k^{\alpha-1}}{2^{\alpha+1}\pi^\alpha R^{\alpha}} (\sin{\Theta})^{-\alpha}e^{i(kR\Theta-\omega t)},
\ee
describes a wave of angular frequency $\omega$ propagating away from the source point at the speed of light. As $\Theta\rightarrow \pi$, $R\Theta=\lvert x-x'\rvert$ approaches the half circumference of $S^d$ and the wave converges at a caustic point antipodal to the source as $D_{\tau-\alpha}^\alpha(\cos{\Theta}-i0)$ diverges, then continues to propagate around $S^d$ and back toward the source for $\Theta>\pi$; there is no actual source or sink at the antipodal point. 

The retardation condition generalizes accordingly, with the requirement that the time-dependent Green function vanish for $kR\Theta_{tot}-\omega t>0$, where $\Theta_{tot}$ includes the cumulative distance from loops around the hypersphere. The presence of incoming as well as outgoing waves at times $t>\pi R/c$ also complicates the imposition of simple outgoing-wave boundary conditions at large distances $R\Theta$ to determine ${\cal G}_R^d$; the causality or retardation condition must be applied directly.

We can rewrite \eq{GRS1} in terms of Gegenbauer functions ${\mathsf D}_{\tau-\alpha}^\alpha(\cos{\Theta})$ and ${\mathsf C}_{\tau-\alpha}^\alpha(\cos{\Theta})$ ``on the cut," analogous to the Ferrers functions or Legendre functions on the cut used by Cohl, Dang, and Dunster in \cite{CohlDangDunster}. These are defined as  \cite{Askey}
\ba
\label{Dcut}
{\mathsf D}_\lambda^\alpha(x) &=& -ie^{-i\pi\alpha}\left(e^{i\pi\alpha}D_\lambda^\alpha(x+i0)-e^{-i\pi \alpha}D_\lambda^\alpha(x-i0)\right), \\
\label{Ccut}
{\mathsf C}_\lambda^\alpha(x) &=&e^{-i\pi\alpha}\left( e^{i\pi\alpha}D_\lambda^\alpha(x+i0)+e^{-\pi i\alpha}D_\lambda^\alpha(x-i0)\right) \\
&=& C_\lambda^\alpha(x\pm i0).
\ea
This gives
\be
\label{GRS3}
{\cal G}_R^d(\Theta,\omega) =  i \frac{\Gamma(\alpha)}{4\pi^{\alpha}R^{2\alpha-1}} \left({\mathsf C}_{\sqrt{k^2R^2+\alpha^2}-\alpha}^{\alpha}(\cos{\Theta})-i{\mathsf D}_{\sqrt{k^2R^2+\alpha^2}-\alpha}^{\alpha}(\cos{\Theta})\right).
\ee

The solution ${\mathfrak G}_{R,\beta}^{d,-}(x,x')$ proposed for the Green function on $S^d$ by Cohl, Dang, and Dunster   in terms of Ferrers functions in \cite{CohlDangDunster}, Eq.~(4.24), is equivalent to the result for the retarded Green function in \eq{GRS2}. This may be shown using \cite{dlmf}, Eq.~14.32.2 and the relation between Legendre and Gegenbauer functions of complex argument in \eq{DQrelation}. However, the result in \eq{GRS2} was derived directly and did not require an appeal to the Euclidean limit to establish its validity and normalization. Their second proposed solution $S_{R,\beta}^{d,-}(x,x')$  (\cite{CohlDangDunster}, Eq.~(4.23)) includes advanced as well as retarded components and describes standing rather than running waves. 

${\cal G}_R^d(\Theta,\omega)$ reduces properly to the the flat-space Green function on $E^d$ for short enough wavelengths and small enough angles  that $kR\Theta\gg 1$ with $\Theta\ll 1$. The difference of the geometries on $S^d$ and on its tangent space at $\Theta=0$ is then negligible. In this limit \cite{LDasymptotics}
\be
\label{SdtoE}
{\mathsf C}_{\nu-\alpha}^\alpha(\cos{\Theta})-i{\mathsf D}_{\nu-\alpha}^\alpha(\cos{\Theta}) \sim \frac{\sqrt{\pi}}{\Gamma(\alpha)}2^{-\alpha+\frac{1}{2}}(kR)^{\alpha-\frac{1}{2}}(\sin{\Theta})^{-\alpha+\frac{1}{2}}H_{\alpha-\frac{1}{2}}^{(1)}(kR\Theta),
\ee
and 
\be
\label{flatGR}
{\cal G}_R^d(\Theta,\omega)
\sim \frac{i}{4}\left(\frac{k}{2\pi R\Theta}\right)^{\alpha-\frac{1}{2}} H_{\alpha-\frac{1}{2}}^{(1)}(kR\Theta), \quad \alpha=\frac{d-1}{2}.
\ee
This is the proper Euclidean limit as noted in \cite{CohlDangDunster}, Eq.~4.9, with $R\Theta=\lvert x-x'\rvert$ the separation of the source and field points.


\subsection{A Mehler-Fock type transform  $\cos{\theta}\in(-1,1)$ \label{subsec:MehlerFock_onSd}}


It is straightforward to show directly that the action of the wave operator on $G_R(\theta,\theta',\omega)$ gives the expected generalized delta distribution in \eq{Gequation}. We first rewrite \eq{wave_eq_S} as
\be
\label{wave_eq_S3}
\left[-\bigtriangleup_\theta-k^2R^2\right]w(\theta)=\left(-\frac{1}{(\sin{\theta})^{2\alpha}}\frac{d}{d\theta}(\sin{\theta})^{2\alpha}\frac{d}{d\theta}-k^2R^2\right)w(\theta)=0
\ee
for $w(\theta)$ a Gegenbauer function. This operator gives zero when acting on $G_R(\theta,\theta',\omega)=G_-(\theta,\theta',\omega)$, \eq{GonSdefined2}, except in the neighborhood of $\theta=\theta'$ where the first derivative increases discontinuously from $\theta<\theta'$ to $\theta>\theta'$  by an amount equal to the Wronskian 
\ba
 {\cal W}(D_{\nu-\alpha}^\alpha,C_{\nu-
 \alpha}^\alpha)_{\theta-i0} &\equiv& D_{\nu-\alpha}^\alpha(\cos{\theta}-i0)\frac{d}{d\theta}C_{\nu-\alpha}^\alpha(\cos{\theta})-C_{\nu-\alpha}^\alpha(\cos{\theta})\frac{d}{d\theta}D_{\nu-\alpha}^\alpha(\cos{\theta} -i0)\nonumber \\
 \label{W(D,C)S}
 &=& -ie^{2\pi i\alpha}\,2^{2\alpha-1}\frac{\Gamma(\nu+\alpha)}{\left[\Gamma(\alpha)\right]^2\Gamma(\nu-\alpha+1)}(\sin{\theta})^{-2\alpha}.
 \ea
The result is a unit step function $\vartheta(\theta-\theta')$ in the action of the first derivative on $G_R(\theta,\theta',\omega)$. The second derivative  then gives  the expected delta distribution $\delta(\theta-\theta')/(\sin{\theta})^{2\alpha}$.  

With this established, we return to the represntation of $G_R(\theta,\theta',\omega)=G_-(\theta,\theta',\omega)$ in \eq{GonSdefined}. When we apply the wave operator $\left(-\bigtriangleup_\theta-k^2R^2\right)$ to this expression we obtain the form of the Mehler-Fock kernel on $S^d$ in \eq{Mehler_kern3} independently of our use of analytic continuation from the hyperbolic case, with
\be
\label{MFS1}
\delta(\theta-\theta')/(\sin{\theta'})^{2\alpha} =  \frac{e^{-2\pi i\alpha}}{2\pi}\int_{C_-}d\nu D_{\nu-\alpha}^\alpha(\cos{\theta_>}-i0)C_{\nu-\alpha}^\alpha(\cos{\theta_<})\frac{2^{2\alpha}\nu\left[\Gamma(\alpha)\right]^2\Gamma(\nu-\alpha+1)}{\Gamma(\nu+\alpha)}. 
\ee
The integration contour $C_-$  on $\nu$ initially runs from $-\infty$ to $\infty$ a small arbitrary distance $a$ above the negative real axis, but can be distorted for $\nu>-\alpha$ to run $+\infty$ either above or below the real axis.

To put the kernel in the usual form of a Mehler-Fock kernel, we will split the integration at $\nu=0$, change $\nu$ to $-\nu$ on the segment $(-\infty,0)$, and then combine the results of the two integrals. This gives
\ba
\delta(\theta-\theta')/(\sin{\theta'})^{2\alpha} &=& \frac{e^{-2\pi i\alpha}}{2\pi}\int_0^{\infty-ia}d\nu\left[D_{\nu-\alpha}^\alpha(\cos{\theta_>})-D_{-\nu-\alpha}^\alpha(\cos{\theta_>})\right]C_{\nu-\alpha}^\alpha(\cos{\theta_<}) \nonumber \\
\label{MFS2}
&&\times \frac{2^{2\alpha}\nu\left[\Gamma(\alpha)\right]^2\Gamma(\nu-\alpha+1)}{\Gamma(\nu+\alpha)} \\
&=& \frac{e^{-i\pi\alpha}}{2\pi}\int_0^{\infty-ia}d\nu \,C_{\nu-\alpha}^\alpha(\cos{\theta})C_{\nu-\alpha}^\alpha(\cos{\theta'})\frac{\sin{\pi\nu}}{\sin{\pi(\nu-\alpha)}} \nonumber \\
\label{MFS3}
&&\times \frac{2^{2\alpha}\nu\left[\Gamma(\alpha)\right]^2\Gamma(\nu-\alpha+1)}{\Gamma(\nu+\alpha)},
\ea
where we have used the relation (\cite{DFS}, Eq.~(3.2))
\be
\label{CDconnection}
C_{\nu-\alpha}^\alpha(z) = e^{-i\pi\alpha}\frac{\sin{\pi(\nu-\alpha)}}{\sin{\pi\nu}}\left[D_{\nu-\alpha}^\alpha(z)-D_{-\nu-\alpha}^\alpha(z)\right]
\ee
to combine the two terms.  It is no longer necessary to distinguish the limits $\theta\gtrless\theta'$ or $\cos{\theta}\pm i0$ in the resulting expression. 

We immediately obtain a generalized Mehler-Fock transform on the interval $0<\theta<\pi$  appropriate for $S^d$
using the expression for the kernel in \eq{MFS3},
\ba				
\label{Mehler_on_S}
{\tilde f}(\nu) &=& \int_0^\pi\frac{C_{\nu-\alpha}^\alpha(\cos{\theta'})}{\sin{\pi(\nu-\alpha)}}f(\theta')(\sin{\theta'})^{2\alpha}d\theta',\\
f(\theta) &=&\label{Mehler_inv_on_S}
 \frac{e^{-i\pi\alpha}}{\pi}\int_0^{\infty-ia}d\nu\,{\tilde f}(\nu)C_{\nu-\alpha}^\alpha(\cos{\theta})\frac{2^{2\alpha-1}\nu\sin{\pi\nu}\left[\Gamma(\alpha)\right]^2\Gamma(\nu-\alpha+1)}{\Gamma(\nu+\alpha)}.
\ea
The result has the same form as that appropriate for the hyperbolic case, Eqs.~(\ref{Mehler_trans})-(\ref{r(lambda,alpha)}) with the expected replacement of hyperbolic by spherical angles and $i\lambda$ by $\nu$.

We note that the function  $C_{\nu-\alpha}^\alpha(\cos{\theta})$ diverges as $(\sin{\theta})^{-2\alpha+1}$ for $\theta\rightarrow\pi$, but that the  integral in \eq{Mehler_on_S} converges  for  $f(\theta')$ finite for $\theta'\rightarrow\pi$ because of the natural integration weight $(\sin{\theta'})^{2\alpha}$. The weight in $\theta$ does not appear in  \eq{Mehler_inv_on_S}; we therefore take  $\theta<\pi$ in that equation.

The Mehler-Fock type transform we have constructed here for use in the hyperspherical rather than hyperbolic context is apparently new. A more direct derivation of the transform and more detailed investigation of its range of validity would be of interest.


\section{Summary and Conclusions \label{sec:conclusions}}


We have constructed the causal or retarded radiation Green functions on the hyperbolic and hyperspherical spaces $H^d$ and $S^d$ using a new method based on generalized Mehler-Fock transformations. This method allows easy implementation of the causality condition and proof of the uniqueness of the soutions. The results clarify and extend those of Cohl, Dang, and Dunster \cite{CohlDangDunster}, and resolve an uncertainty in their proposed solutions of the problem. Our results hold for general values of the dimension $d$, which need not be integer or real.

Our method made extensive use of the kernel of the combined Mehler-Fock transform and its inverse. This is a Schwarz distribution which has the form of the  source term for radiation in the inhomogeneous wave equation. The necessary Mehler-Fock transform was known for the case of radiation on $H^d$. The derivation of corresponding results for $S^d$ was initially accomplished by an analytic continuation of the kernel appropriate to the Helmholtz equation on $H^d$, an example of the continuation of a distribution. This was used the construct the retarded radiation Green function on $S^d$. This was shown to be correct and unique, and in turn allowed the proof of an apparently new form of the Mehler-Fock transform applicable for spherical angles $\theta$ on the interval $(0,\pi)$.

 
\begin{acknowledgments}
The author would like to thank the Aspen Center for Physics, which is supported by The National Science Foundation grant  PHY-1607611, for its hospitality while parts of this work were done. 
\end{acknowledgments}


\section*{Author declarations}

\section*{Conflict of Interest}
The author has no conflicts of interest with respect to this work.

\section*{Data availability}
Data sharing is not applicable to this article as no new data were created or analyzed in this study.



\begin{thebibliography}{99}
\bibitem{outgoing_waveBC}
An alternative approach to identifying the Green function uses the imposition of outgoing-wave boundary conditions to ensure that  ${\cal G}_R^d(\Theta,\omega)$ describes emission from a source. The function $D_{-i\sigma-\alpha}^\alpha(\cosh{\Theta})$ is proportional to $e^{ikR\Theta}$ for $kR\Theta\rightarrow\infty$, and satisfies that condition; the remaining independent solutions all involve incoming-wave components proportional to $e^{-ikR\Theta}$ for $kR\Theta\rightarrow\infty$.
\bibitem{footnote2}
The flat-space limit was examined in detail in \cite{CohlDangDunster}. However, those authors took the  limit as $R\rightarrow\infty$ with $\omega$ and $\Theta$ fixed rather than the physical short-wavelength limit for wave propagation on a fixed hyperboloid of dimension $d$ and hyperbolic radius $R$. This reduction was the basis on which they identified the proper solution of the wave equation to use in the Green function on $H^d$, and  to determine the  normalization of that Green function. They did not, however, recognize the uniqueness of the result. The causality condition did not enter in their argument, but had been used indirectly in the Euclidean case through the imposition of outgoing-wave boundary conditions on the Green function in $E^d$; this eliminates incoming waves at $R\rightarrow\infty$. This method did not yield a unique result for the Green function on $S^d$, but the proper solution can be determined by using the causality condition directly.
\bibitem{footnote3}
For a general discussion of the continuation of distributions to forms expressible  as the difference of boundary values of analytic functions see \protect{\cite{BremermannDurand}.} We do not explicitly need  those results  or generalizations of them here as the kernel distributions are already expressed directly in terms of the Dirac delta distribution multiplied by functions of $\theta$ and $\theta'$, or as  integrals of analytic functions of those variables. However, regularization of the delta distribution or of the divergent integrals leads to results of the type discussed there.
\end{thebibliography}

\end{document}